# Methodology for Jointly Assessing Myocardial Infarct Extent and Regional Contraction in 3D-CMRI

Y. Chenoune*, *Member*, *IEEE*, C. Pellot-Barakat, C. Constantinides, R. El Berbari, M. Lefort,
E. Roullot, E. Mousseaux and F. Frouin, *Member, IEEE*

*Abstract*—
**Automated extraction of quantitative parameters from Cardiac Magnetic Resonance Images (CMRI) is crucial for the management of patients with myocardial infarct. This work proposes a post-processing procedure to jointly analyze Cine and Delayed-Enhanced (DE) acquisitions in order to provide an automatic quantification of myocardial contraction and enhancement parameters and a study of their relationship. For that purpose, the following processes are performed: 1) DE/Cine temporal synchronization and 3D scan alignment, 2) 3D DE/Cine rigid registration in a region about the heart, 3) segmentation of the myocardium on Cine MRI and superimposition of the epicardial and endocardial contours on the DE images, 4) quantification of the Myocardial Infarct Extent (MIE), 5) study of the regional contractile function using a new index, the Amplitude to Time Ratio (ATR). The whole procedure was applied to 10 patients with clinically proven myocardial infarction. The comparison between the MIE and the visually assessed regional function scores demonstrated that the MIE is highly related to the severity of the wall motion abnormality. In addition, it was shown that the newly developed regional myocardial contraction parameter (ATR) decreases significantly in delayed enhanced regions. This largely automated approach enables a combined study of regional MIE and left ventricular function.**

*Index Terms*— **Cardiac MRI, Contraction Function, Delayed-Enhancement Quantification, Registration, Segmentation.**

## I. INTRODUCTION

QUANTITATIVE analysis of cardiac images is essential for the diagnosis and follow-up of ischemic heart diseases. Many clinically relevant parameters can be extracted from cardiac images, acquired with various imaging modalities. Among these modalities, Cardiac Magnetic Resonance Imaging (CMRI) is a minimally invasive technique providing anatomical and functional images, both suitable for ischemia detection, viability assessment and ventricular function study.

Our objective is to help clinicians save data analysis time, by providing software tools allowing the estimation of cardiac functional parameters, in an automated and reproducible way. In the clinical routine, generic tools are often required for the left ventricular (LV) borders segmentation and for CMR images registration. Some dedicated methods for the automatic estimation of cardiac quantitative parameters can also be very useful to help the clinicians in the diagnosis and therapeutic decision making.

Delayed-Enhancement CMRI (DE-CMRI) is a recent technique which has proven its capability to identify and quantify myocardial infarction. Several studies have shown the ability of DE-CMRI to distinguish permanently dysfunctional myocardium from viable segments, able to recover contractile function after revascularization [1]. Indeed, delayed images obtained less than 10 minutes after Gadolinium-based contrast agent injection demonstrate local hyperenhancement, corresponding to myocardial necrosis.

Complementarily to DE-CMRI, the analysis of Cine-CMRI allows the qualitative and quantitative evaluation of cardiac motion and ventricular function. Both visual and automated analyses of amplitude and temporal changes in the wall motion allow the detection of regional alterations in the ventricular function.

The analysis of both DE and Cine CMRI is thus essential for the diagnosis and follow-up of ischemic heart disease. Several authors have studied the late gadolinium enhancement and the myocardial contraction for the prediction of myocardial function recovery after a recent infarct event or after revascularization [2-6].

In this paper, we propose to study the enhancement and myocardial contraction jointly in the case of already diagnosed infarct. The primary aim of our study is to propose a complete

Manuscript submitted November 20, 2011. This work was supported by the *UPMC/Inserm U678 research unit*, 91 Boulevard de l'Hôpital, 75634, Paris, France and by the *ESME-Sudria Engineering School*, 38 rue Molière, 94200, Ivry-sur-Seine, France. Asterisk indicates corresponding author.

*Y. Chenoune and C. Constantinides are with the *UPMC/Inserm U678 research unit*, 91 Boulevard de l'Hôpital, 75634 Paris, France and with the *ESME-Sudria Engineering School*, 38 rue Molière, 94200, Ivry-sur-Seine, France (chenoune@esme.fr, constantin@esme.fr).

C. Pellot-Barakat, M. Lefort, and F. Frouin are with the *UPMC/Inserm U678 research unit*, 91 Boulevard de l'Hôpital, 75634, Paris, France (claire.barakat@imed.jussieu.fr, mlefort@imed.jussieu.fr, frouin@imed.jussieu.fr).

R. El Berbari was with the *UPMC/Inserm U678 research unit*, she is now with the *Faculty of Engineering, ECCE Department at Notre Dame University*, 5725 Deir-El-Kamar, Lebanon (rberbari@ndu.edu.lb).

E. Roullot is with the *ESME-Sudria Engineering School*, 38 rue Molière, 94200 Ivry-sur-Seine, France (roullot@esme.fr).

E. Mousseaux is with the *UPMC/Inserm U678 research unit* and with the *APHP, Hôpital Européen Georges Pompidou*, Cardiovascular Radiology Department, 20 rue Leblanc 75908, Paris France (elie.mousseaux@imed.jussieu.fr).

> SUBMISSION TO IEEE TRANSACTIONS ON BIOMEDICAL ENGINEERING – APRIL 2012 – YASMINA CHENOUNE 2Cine/DE CMRI processing chain in order to provide an automated quantification of the Myocardial Infarct Extent (MIE) and to study the myocardial contraction. The second goal is to determine the relationship between the regional hyperenhancement and the associated regional myocardial contraction, which is a known complex problem [2].

Usually, the assessment of the transmural extent of the infarction from DE-images is performed visually [8] but several computer-assisted methods have been proposed to automatically or semi-automatically detect and quantify the segmental MIE [5, 9-12]. However, most of these methods require the manual delineation of the myocardial borders, which can be very fastidious and time consuming in the clinical routine.

Although many segmentation methods such as deformable meshes [13], level sets [14-17], diffusion wavelets and boosting [18], graph-cuts [19] or intensity gradient and texture features [20, 21] have been proposed to automatically or semi-automatically segment the LV cavity from Cine-CMRI, fewer works have dealt with the segmentation of DE-CMRI. Indeed, Cine-CMR images present a good contrast between myocardium and cavities that allows an accurate delineation of the myocardium borders, while DE images are noisy and enhanced in infarcted zones which makes the detection of the endocardial contours difficult. Based on that idea, Dikici et al. used a prior model of the myocardial borders extracted from Cine-CMR images as well as image features present in the DE images to segment the myocardium [22]. Another group recently proposed an automatic method using geometrical template deformation and LV cavity shape prior [23] to directly extract the myocardial contours from DE images.

In the following study, the DE image segmentation is indirectly performed by superimposing the myocardial contours extracted from the highly contrasted anatomical Cine images onto the corresponding DE images. This process requires the prior registration of Cine/DE images to correctly align these two kinds of images. There are several works related to the registration of different MR images such as DE-CMRI and perfusion images for the visualization of infarction [24], but not many studies are related to the DE/Cine CMRI registration. In that framework, a method applied to the moving propagation of suspicious myocardial infarction was recently published [25].

In the work presented here, a coarse spatiotemporal alignment is first achieved to correct for temporal and global scan misalignment [26]. To correct for local misalignments, due to patient and/or respiratory movement artifacts, a 3D rigid registration method is then applied. Using previously obtained results, dedicated procedures are subsequently proposed to automatically quantify the MIE and to study the regional LV function. A new meaningful contraction index is proposed to characterize and quantify the regional myocardial contraction. The relationship between the regional hyperenhancement and the associated regional myocardial contraction is then studied.

In the following sections, the complete procedure for analysis of CMRI is presented. Section II describes the acquisition protocol and studied data. Section III details the generic processing of CMRI, including a two-steps procedure for multimodal data registration and a semi-automated segmentation method for the LV myocardial contours extraction. Section IV exposes the quantification approaches developed for the assessment of the MIE and myocardial contraction. Finally, the results of the registration and segmentation steps, the quantification of the MIE, the contraction study as well as the relationship between enhancement and contraction are reported and discussed in Section V.

## II. CMRI DATA AND PROTOCOL

### A. Data acquisition

We studied 10 patients (2 women, 8 men; mean age 59±19 years) with clinically proven myocardial infarction. Institutional review board approval was obtained from all the patients included in the study. All patients underwent 3D+t Cine-CMRI exams (25 to 35 phases) with electrocardiographic (ECG) gating on a 1.5 T MR-scanner (GE Medical Systems). The contraction sequences were obtained with breath-holding using Fast Imaging Employing STeady-state Acquisition (FIESTA), in 12 to 15 short-axis (SA) views. The imaging parameters were: a repetition time of 3.5 to 4 ms, an echo time of 1.6 to 1.7 ms, a flip angle of 50 degrees, an image matrix of 512x512, a slice thickness and a spacing between slices of 8 mm and a pixel size varying from 0.72x0.72 $mm^2$ to 0.86x0.86 $mm^2$.

Three dimensional DE-CMRI were then acquired with breath-holding, during the diastolic phase to minimize cardiac motion, using standard segmented inversion-recovery sequences at 5, 6 and 7 minutes after the injection of Gd-DTPA (0.2 mmol/Kg body weight). The heart was covered in 24 to 44 SA views. The imaging parameters included a repetition time of 4.3 to 5.5 ms, an echo time of 1 to 2 ms, a flip angle of 15 degrees, an inversion time of 150 to 200 ms, a trigger delay time for inversion recovery pulse of 300 ms, an image matrix of 256x256, a slice thickness of 6 to 7 mm, a spacing between slices of 3 to 3.5 mm and a pixel size varying from 1.36x1.36 $mm^2$ to 1.48x1.48 $mm^2$.

### B. Myocardial Segment Definition

For the regional myocardial function study, the LV cavity was divided into 6 equiangular circumferential segments (A: anterior, AL: anterolateral, IL: inferolateral, I: inferior, IS: inferoseptal and AS: anteroseptal), according to the standardized LV model [27], on the Cine SA views (mean of 6 slices/patient). The origin of the division was defined at the anterior insertion of the right ventricle into the interventricular septum and the segments were defined in a clockwise manner, as shown in Fig. 1(c).

For a precise regional analysis of the myocardial enhancement, considering the hyperenhancement heterogeneity in the myocardial segments, the 6 previously defined segments were divided into 3 sub-segments, as previously proposed in [11]. This resulted for each SA slice in 18 sub-segments as shown in Fig. 1(c).

### III. GENERIC PROCESSING

Fig. 1 illustrates the complete procedure workflow. The selection of Cine and DE-CMRI data are presented in Fig. 1(a). The generic processing including the segmentation and the registration steps are illustrated in Fig. 1(b). The strategy is to exploit the good contrast of the anatomical Cine images to extract the myocardial contours. These contours can then be superimposed on the corresponding DE images to assist the MIE quantification. The application of dedicated processing to the MIE quantification and the extraction of meaningful parameters to study the contraction as well as the relationship between hyperenhancement and contraction are indicated in Fig. 1(c).

#### A. Temporal Synchronization and Scan Alignment

To compensate for the differences in the acquisition parameters between the DE and the Cine images (different temporal resolutions, various voxel sizes), *a posteriori* temporal synchronization and scan alignment of the DE/Cine datasets were first carried out.
A 3D averaged Cine image was first calculated in the temporal zone, corresponding to the DE volume acquisition phase of which duration was about 130 ms. This corresponded to 4 or 5 phases of the Cine sequence. The spatial scan alignment between the two datasets was then performed to adjust scales and voxel sizes, using the image orientation and position information [26].

#### B. DE/Cine 3D-CMRI Refined Rigid Registration

After the scan alignment, a second rigid registration step was necessary to finely register the DE/Cine datasets about the heart region. Although the images were acquired during the diastolic phase where the heart motion is limited, and even though the temporal synchronization minimized the cardiac motion effect, there were still residual misalignments mainly due to patient motion and possible differences in breath-holding positions.

As the spacing between slices is about 2.3 to 2.7 times smaller in the DE volume (3 to 3.5 mm) than in the Cine volume (8 mm), the slices among the DE available images that best match the Cine slices have to be determined. For that, the optimal transformation $T_{opt}$ that matches the entire DE volume to the Cine one is estimated. Since a preliminary global alignment has been performed, the field of search of the translation $\Delta z$ in the longitudinal direction can be limited to an interval of ±1 pixel (displacement of ±3 to ±3.5 mm). Besides, tests showed that increasing the field of search for $\Delta z$ shifts the slices to another anatomical slice level. To finely register the DE/Cine datasets, sub-volumes $I$ and $J$ constituted by the heart zone regions including the right ventricle (RV) are extracted and matched. The displacement between the two sub-volumes is modeled by a 2D rigid transformation in the xy plan that was found optimal and estimated using the Normalized Mutual Information (NMI) as a similarity measure [28-29]. The sub-volumes are considered well matched when the value of the NMI is maximal. The Powell optimization method [30] is used to find the rigid transform $T$ (rotation and translations) minimizing the cost function:

$$C(T) = e^{-NMI(I,J,T)} \qquad (1)$$

The obtained transformation $T_{opt}$ was then used to transform the whole DE volume to map the corresponding Cine one. The tri-linear interpolation was used to estimate the gray level values of the new pixel positions.

#### C. Semi-Automated Myocardial Segmentation

For the contractile function study, the endocardial contours were obtained from the end-diastolic 3D Cine image, where the myocardium reaches its maximal relaxation. It provides the outer border, needed for the contraction study using the Parametric Analysis of Main Motion (PAMM) (Section IV.B). For the enhancement study, the LV epicardial and endocardial (epi/endo) contours were extracted from the 3D averaged Cine image, previously seen in Section III.A, which is more representative of the DE image acquisition period than the end-diastolic one (see Fig.1 for illustration)."

The segmentation [31] was semi-automatically achieved using a filtering combining morphological openings and closings [32], and applying the Gradient Vector Flow Snakes (GVF-Snakes) method [33] to the filtered images.

*Endocardial Border Segmentation*

For each SA slice, the user defined two points on the image: $P_0$ in the LV cavity and $P_1$ at the anterior insertion of the right ventricle into the interventricular septum. A squared ROI was then automatically centered on $P_0$ and its side dimension was set to three times the distance between $P_0$ and $P_1$. Morphological filters combining openings and closings on connected sets were then applied to the ROI. A size parameter $\lambda$ was defined such that the filtered image contains only connected components of area larger than $\lambda$ [32]. Filtered images with homogeneous regions and in which the contrast between the papillary muscles and the LV cavity was reduced were obtained. Varying the size parameter $\lambda$ from 5% to 80% of the ROI surface allowed obtaining a set of filtered images.

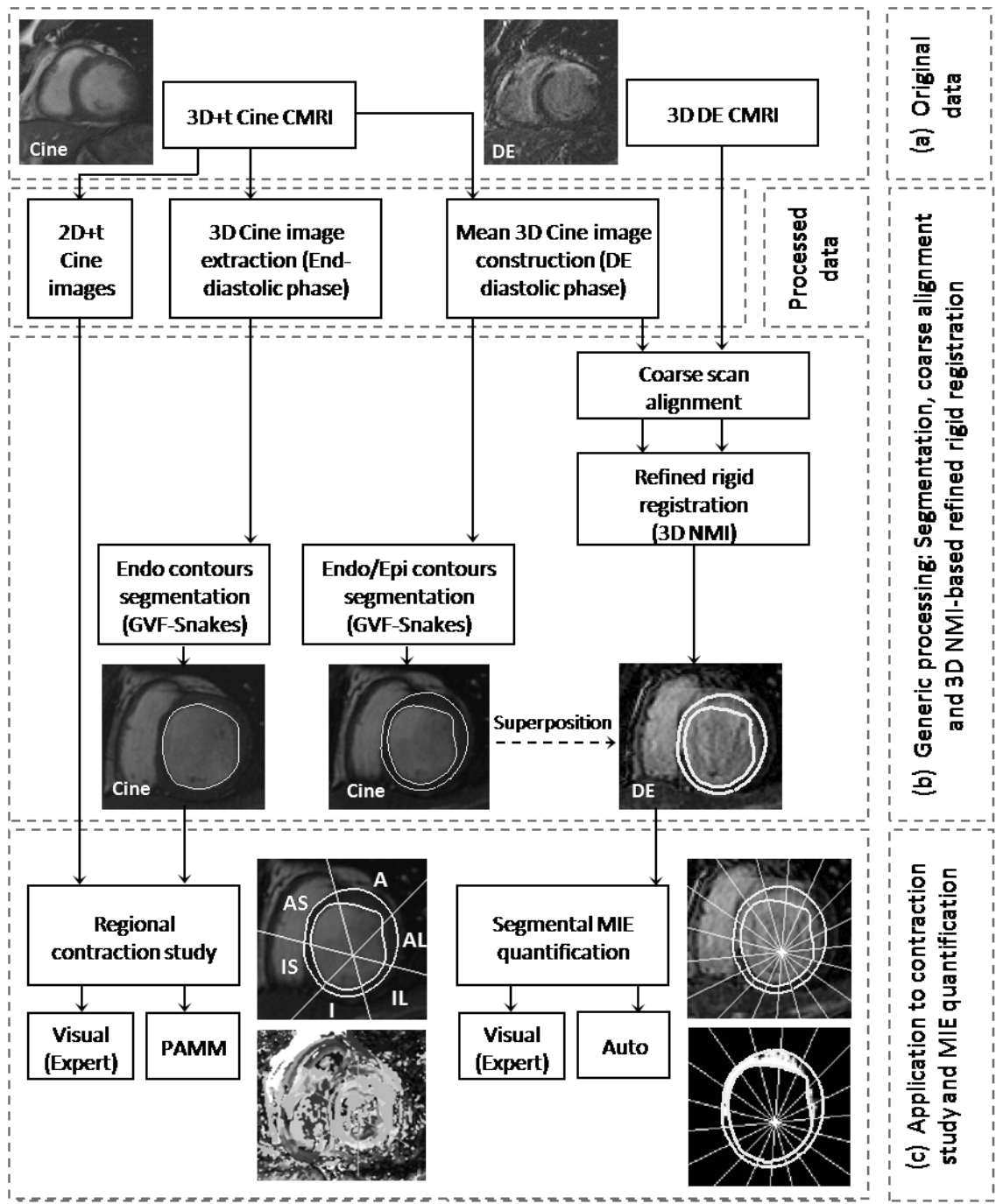

Fig 1. The complete procedure workflow, showing (a) the selection of Cine and DE-CMRI data; (b) the generic processing step including the segmentation of epicardial (epi) and endocardial (endo) contours, and the DE/Cine-CMRI coarse and refined 3D registration; (c) the application of the automatic and visual analyses to the regional contraction study and to the MIE quantification.

The ratio between the filtered surface including $P_0$ and $\lambda$ was computed for all the filtered images and the image giving the ratio closest to 1 was defined as the best filtered image [34]. The user could refine this value of $\lambda$ according to the appearance of the filtered image. The next step was the segmentation of the endocardial borders using $P_0$ as an initialization of the GVF-Snake. Weighting parameters used to control the snake evolution in the spatial domain were: $\alpha = 1$, (elasticity), β=40 (rigidity), $\kappa_p = 0.6$ (pressure force), $\kappa = 1.7$ (GVF force) and $\mu_{gvf} = 0.3$ (GVF regularization parameter) [31].

*Epicardial Border Segmentation*

The low epicardium-background contrast makes epicardial borders difficult to segment. Consequently, a restriction of the GVF-Snake evolution in a restrained area was proposed. The internal limit of this mask was obtained by widening the previously segmented endocardial contour by 2 or 3 pixels. The external limit was obtained by widening the endocardial contour with a distance close to the anatomical thickness of the myocardium in the diastolic phase (7 to 12 mm). The values of the internal and external limits of the mask were adjustable by the user. The initial position of the contour was defined within the limits of the mask and the GVF-Snake parameters were similar to those used for the endocardial border segmentation. However pressure forces ($\kappa_p$) decrease through the iterative process, thus avoiding unnecessary overexpansion of the snake.

## IV. MIE QUANTIFICATION AND REGIONAL FUNCTION STUDY

This section presents the dedicated processing that was developed for the MIE quantification and the myocardial contraction study. The guidelines used for the visual expertise are then explained. The methods of statistical analysis are finally given.

### A. MIE Automated Assessment

The computer-assisted method using a Fuzzy c-means algorithm and previously proposed in [11], was applied to all the DE exams to automatically detect and quantify the segmental MIE. On each SA slice, the myocardium was first divided into four concentric layers, from the endocardial border to the epicardial border. These layers were used in the scoring of the infarct which spreads in a wavefront extension from the endocardium to the epicardium. The 18 previously defined myocardial sub-segments (Section II.B) were then automatically categorized according to a 5-point scale in which a score of 0 indicated no infarction and a score of 1 to 4 indicated a hyperenhancement extent respectively: inferior to 25%, from 26% to 50%, from 51% to 75% and from 76% to 100%. A segment was considered as transmural if it was graded 3 or 4. Since myocardial infarcts initially occur in the subendocardial muscle and may then extend to the epicardium in the case of transmural infarcts, segmental epicardial scar cannot occur without the associated presence of endocardial scar. Thus, the algorithm analyzes the layers sequentially from the endocardium towards the epicardium and stops as soon as a healthy layer is found. The default assigned score is 0. The score is incremented by 1 every time an enhanced layer is found. As no significant differences in the results for the 5, 6 and 7 min DE studies had been reported in a previous paper [35], combined sub-segmental indexes were finally defined from the three DE studies using a majority rule.

### B. Regional Myocardial Function Study

The local contractile function was quantitatively analyzed on the Cine-CMRI, using PAMM [36]. This method allows synthesizing the information contained in a sequence corresponding to one cardiac cycle, in parametric images related to the contraction amplitude and timing. For that purpose, the intensity-time curve $P(x, y, t)$ of each pixel $(x, y)$ was observed during the cardiac cycle. Considering a pixel belonging to the cavity and close to the endocardial contour during the diastolic phase, its intensity will decrease from that of the LV cavity to that of the myocardium during the systolic phase. The intensity-time curve $P(x, y, t)$ was thus modeled using the Window function $g(t)$ as follows:

$$P(x,y,t) = A_b(x,y) - A_v(x,y).\,g\big(t, T_{on}(x,y), T_{off}(x,y)\big) + e(x,y,t) \quad (2)$$

where $A_b(x,y)$ is the signal intensity at end-diastole, $T_{on}(x,y)$ the contraction beginning time at which the pixel $(x,y)$ shifts from the LV cavity to the myocardium, $T_{off}(x,y)$ the relaxation beginning time at which the pixel $(x,y)$ shifts from the myocardium to the LV cavity, $A_v(x,y)$ the signal intensity variation during the contraction phase and $e(x,y,t)$ the error term. The adaptive Window function $g(t)$, defined as follows:

$$\begin{cases} g(t) = 1 & \text{if } t \in [T_{on}(x,y), T_{off}(x,y)] \\ g(t) = 0 & \text{otherwise} \end{cases} \quad (3)$$

was found to provide fast, robust, and precise results, compared to sinusoidal models or data-driven models [36].

Several parameters can be extracted from the amplitude and time parametric images to study the myocardial wall motion. To reduce the noise effects, each parameter was computed inside the six segmental regions delimited by the end-diastolic segmented endocardial contours. A discriminative index, significantly correlated with the wall motion condition was defined. First, a quantitative segmental index was estimated from the parametric image of amplitude $A_v$ and controlled by $A_b$. This index related to the amplitude of displacement of the pixels in the myocardium decreases with the severity of the wall motion abnormality. Therefore, it is high for normal segments and gets lower as the severity of the wall motion pathology increases. Moreover, the mean transition time was defined as the average between the time parameters $T_{on}$ and $T_{off}$, normalized by the cardiac cycle duration [31]. Contrarily to the amplitude, this index related to the time of displacement increases with the severity of the contraction abnormality [37].

Since the amplitude and time parameters act complementarily, a new index combining both parameters, was defined as the ratio between the amplitude and the mean transition time. This combined index, called ATR for Amplitude to Time Ratio, provides a powerful way to characterize the segmental myocardial contraction since it naturally decreases as the severity of the contraction dysfunction increases. The combination of amplitude and time parameters has already been proposed for echocardiographic images and has shown its ability to discriminate between normal and pathological segments [37].

### C. Visual Assessment

A previous study had shown that whether the DE data were acquired 5, 6 or 7 min after injection of the contrast agent, the quantification results were globally identical [35]. Thus, for each patient, the last DE study was arbitrarily chosen for visualization by the expert to validate the MIE automatic assessment. A visual transmurality index, varying from 0 to 4 was attributed to each of the 18 myocardial sub-segments by the expert, according to the scale described for the automatic MIE quantification (Section IV.A). A segment-to-segment comparison of the agreement between the automated and visual segmental MIE scores was performed.

All the Cine sequences were also visually analyzed by the expert to characterize the regional myocardial contraction. For each of the 6 previously defined segments, the myocardial function was described as normokinetic (N), hypokinetic (H) or akinetic/dyskinetic (AD). The automatically and visually assessed parameters were difficult to obtain on the most basal and apical slices, due to the partial-volume effect. These slices were consequently excluded from the analysis.

*D. Statistical Analysis*

One-way ANalysis Of VAriance (ANOVA) was performed with the JMP software (SAS Institute, Cary, NC) to complete different statistical comparisons between the visual and the automatic MIE quantification scores, as well as the ATR and the contraction visual analysis. Results were presented as $mean \pm standard\ deviation$ and were considered as statistically significant when the associated $p$ value was less than 0.05.

## V. RESULTS AND DISCUSSION

*A. DE/Cine Rigid Registration*

The DE/Cine CMRI refined registration was performed for the 3 DE exams of all patients, resulting in a total of 30 processed DE volumes, corresponding to 174 DE SA slices. For each patient, rotation and translation parameters in the ($x,y,z$) directions were estimated. The analysis of the values of the optimal transformations showed a mean absolute displacement value of 2.18±1.76 mm in the $x$ direction, 2.37±2.04 mm in the $y$ direction and 1.75±1.56 mm in the $z$ direction. The mean absolute value of the rotation $\theta$ in the ($x,y$) plane was 0.73±0.86. A qualitative evaluation of the 3D DE/Cine CMRI registration quality was carried out on all the DE datasets. To achieve it, an experienced physician counted the number of correctly superimposed epicardial and endocardial contours on the DE images before and after the 3D refined registration. The improvement was systematic in the initially misaligned data, corresponding to 22 of the 30 DE exams. High-quality alignment results were observed in more than 92% of the slices of the registered datasets.

To obtain accurate registration results, different choices related to the ROI definition, the histogram computing and the optimization process were made. To extract the sub-volumes to register, a ROI including all the heart structure was defined on a Cine median slice. It is important to include the right ventricle (RV) as it shows similar intensities in the cine and DE images which can compensate for the possible intensity inversion that can occur in the enhanced LV of DE images and lead to matching uncertainties. In practice, we exploited the ROI defined for the segmentation and enlarged it by a factor of 2 to 2.8. This ensured a good estimation of the probability distribution, used thereafter for the NMI calculation. Indeed, too few overlapping parts between the images, i.e. too few samples, led to a weak statistical power for the probability distribution estimation. In addition, the mutual information can increase as the misregistration increases, depending on the relative sizes of background and cardiac structures [28]. To generate the joint histogram, we chose a small number of bins (100), for greater noise resistance [38].

An accurate study of all the datasets was performed to know if large, enhanced areas were at the origin of the 8% low-quality alignment between the DE and the Cine images. In most cases, the observed registration errors occurred in the one or two most apical slices whether on largely or mildly infarcted cases. This could be due to the fact that the Cine data were acquired during different breath holds. To correct for these misalignments, an automatic repositioning of the different breath hold sets can be implemented using an automatic approach, such as the one proposed by Elen et al [39]. However, newly available MRI sequences should allow the acquisition of the whole cine volume in one breath hold. In the remaining cases, the quantification errors observed did not result from a poor registration but from an inaccurate segmentation due to noisy and poorly contrasted images. New MRI sequences should also improve the quality of 3D DE images and partly solve this problem.

*B. Myocardial Segmentation*

The segmentation method described in Section III.C was used to semi-automatically define the epicardial and endocardial contours from the 3D averaged Cine-CMRI for the MIE assessment. It was also used to estimate the endocardial contours from the end-diastolic Cine-CMRI for the contractile function study. This corresponded to a total of 58 segmented SA slices, repeated for each case.

The quality of the obtained contours was visually assessed. To overcome the low contrast problem between the left ventricle and the background, a pre-processing step using morphological filters with different size values $\lambda$ was first performed. An adjustment of the size parameter $\lambda$ was necessary in 20 to 25% of the cases to find the filtered image giving the best segmentation result. For these cases, about 3 or 4 values of $\lambda$ were tested to select the best endocardial segmentation (Fig. 2).

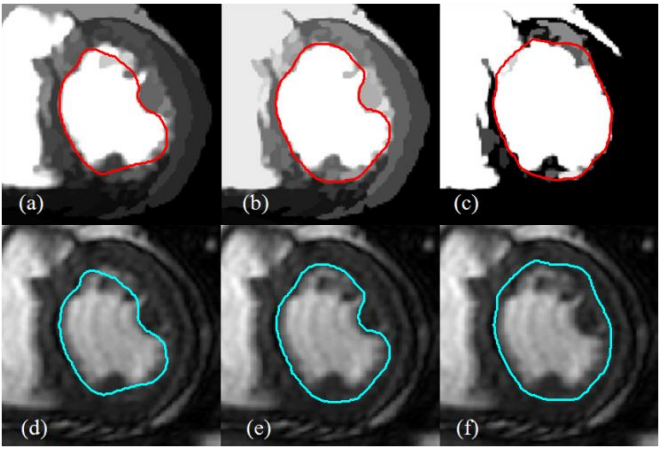

Fig. 2. Filtered Cine-CMR Images and the extracted contours using the GVF-Snakes method for different values of $\lambda$ : (a) $\lambda = 1770$, (b) $\lambda = 2160$ and (c) $\lambda = 5514$, (d-f) superimposition of the contours on the original Cine image, (f) final endocardial segmentation.

The initialization of the GVF-Snake proved to be robust to the position of the $P_0$ point in the LV cavity. The values of the parameters which control the global evolution of the Snake allowed reaching a satisfactory solution in most cases. The segmentation of the apical slices was nevertheless difficult because the contour was quickly attracted beyond the expected endocardial borders. The solution was to decrease the pressure forces γ or to define a smaller ROI, closer to the LV cavity borders. The segmentation of the median slices where papillary muscles are visible in the LV cavity was successfully achieved in the majority of cases. To overcome some encountered difficulties, the user could apply a smooth convex hull around the initially obtained contour. For the segmentation of the epicardium, the user could modify the mask limits. As the GVF-Snake evolution was restrained in the area delimited by the mask, this process was globally faster than the segmentation of the endocardium. Similar epicardial and endocardial contours could also be achieved by incorporating a shape prior to exclude the papillary muscles as in [17].

*C. MIE Quantification*

Figure 3 illustrates the process of MIE quantification with data from a patient with a highly enhanced myocardium (top row) and a weakly enhanced myocardium (bottom row). The myocardial contours obtained for the SA Cine slices of the two patients are shown in Figs. 3(a) and 3(d), their superimposition on the registered corresponding DE images in Figs. 3(b) and 3(e), and the estimation of the enhanced myocardium after the fuzzy c-means algorithm in Figs. 3(c) and 3(f).

Quantitative values of the MIE were automatically obtained for the 10 patients, on the 1044 myocardial sub-segments. It has been shown previously [11] that the inter-observer variability in the analysis of transmural extent is significantly reduced with the subdivision of each slice into 18 sub-segments, instead of six segments. This subdivision makes the transmurality assessment more precise and relevant.

These values were estimated before and after the refined 3D registration step.

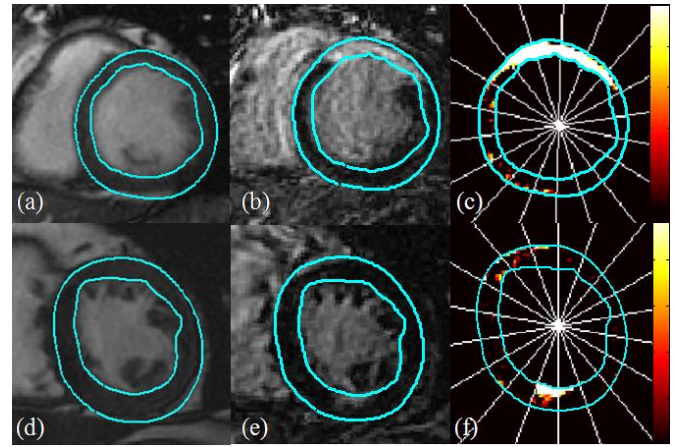

Fig. 3. Illustration of the epicardium and endocardium segmentation on a median slice, from a patient with a highly enhanced myocardium (top row) and a weakly enhanced myocardium (bottom row), (a) and (d) Cine images and segmented epicardial and endocardial contours, (b) and (e) DE corresponding images after registration with the cine-obtained contours superimposition, and (c) and (f) illustration of the Fuzzy c-means MIE quantification results on the 18 myocardial sub-segments.

Fig. 4 illustrates the effect of the refined registration step showing the superimposition of the myocardial contours on the non-locally registered (Fig. 4(a)) and the locally registered corresponding DE images (Fig. 4(c)). The quantification errors in the segments pointed out by arrows in Fig. 4(b) and due to the cavity inclusion in the myocardium segmentation were corrected after the 3D registration, as shown in Fig. 4(d). Ischemic and healthy myocardium zones were thus more precisely identified and quantified. In this example, the sub-epicardial enhancement on the septal size is not classified as scar, since the algorithm stopped at the first layer, as explained in Section IV.A. The classification algorithm is thus able to eliminate possible remaining errors such as sub-epicardial enhancement. For this slice, we observed an absolute agreement with the visual expertise of 72% before and 83% after the refined registration and an agreement with a tolerance of one grade of difference of 78% before and 100% after the registration.

To further evaluate the contribution of the 3D refined registration in the MIE quantification, the automatically obtained MIE scores before and after the refined registration step were compared to the visually segmental grades. Absolute agreements (identical gradation) of 76% and 80% were observed respectively before and after the refined 3D registration. An agreement with a tolerance of one grade of difference, which is tolerated in practice, was found in 86% and 91% of the cases respectively before and after the registration.

Table I gives the table of comparison between the automatic and visual sub-segmental transmurality scores obtained after registration in the 1044 matched sub-segments. A good agreement of the automatically quantified MIE and the expert scores is observed.

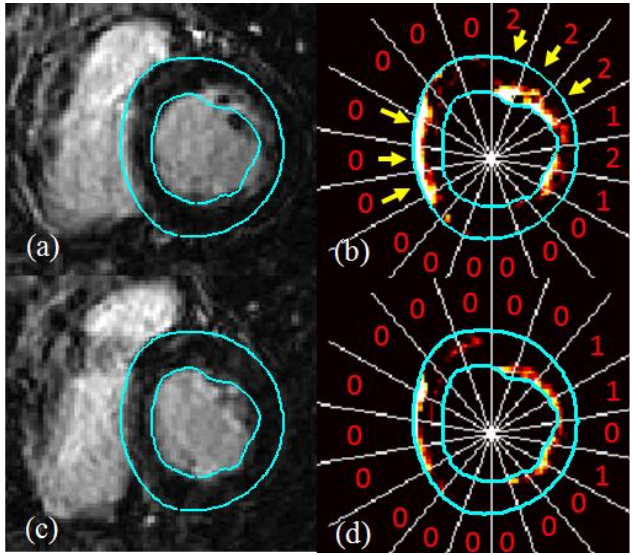

Fig. 4. Example of registration and quantification results, obtained on a basal SA slice: superimposition of the extracted myocardial borders from the Cine-CMRI on the DE corresponding image (a) before and (c) after the 3D refined registration, results of the automatic MIE quantification using the Fuzzy c-means algorithm (b) before and (d) after the 3D refined registration.

A total of 94 sub-segments with a disagreement between automatic and visual MIE assessment of 2 grades or more were identified. A fine analysis of Table I results showed that most differences between automatic and visual scores were observed at the frontier between healthy and infarcted myocardium or between two infarcted zones with different extent values (41 sub-segments). An under-estimation of the infarct extent was observed in 36 sub-segments because of segmentation or registration errors on some slices with poor contrast or residual temporal misalignment. This essentially concerned the apical slices.

TABLE I
AUTOMATIC AND VISUAL SUB-SEGMENTAL TRANSMURALITY SCORES CONCORDANCE AFTER THE 3D REFINED REGISTRATION

| Visual | Automatic segmental indexes | | | | |
| --- | --- | --- | --- | --- | --- |
|  | 0 | 1 | 2 | 3 | 4 |
| 0 | 761 | 26 | 12 | 2 | 4 |
| 1 | 39 | 6 | 5 | 1 | 3 |
| 2 | 27 | 5 | 25 | 1 | 7 |
| 3 | 5 | 3 | 14 | 1 | 19 |
| 4 | 7 | 3 | 20 | 2 | 46 |

Finally, the automatic quantification failed in 17 sub-segments because of poor image quality, due to MRI artifacts or noise. Since the validation by visual assessment is subject to inter-observer variability, a comparison of the MIE score estimated by two independent readers was performed. A total of 86% of the sub-segments were similarly classified (within the same class or the neighboring class). The discordances mainly occurred for high level of MIE where the evaluation task is difficult especially for the sub segments where the myocardial wall is thin. The inter-observer variability is slightly deteriorated compared to that found in [11], probably because the 3D acquired DE images are noisier than the 2D DE images that were used in [11]. However, since inter-observer discordances only concern high MIE cases, we can confidently consider the visual assessment used in this study reliable.

*D. Contraction Study*

The visual analysis of the regional contractile function on the standardized segments (see Section II.B) of the Cine-CMRI was carried out for a total of 348 segments. Among them, 173 were classified as normokinetic, 93 as hypokinetic, and 82 as akinetic or dyskinetic. Fig. 5 illustrates the contraction study on Cine-MR images using the PAMM method. Fig. 5(a) shows the superimposition of the $A_b(x, y)$ and $A_v(x, y)$ parametric images and Fig. 5(b) shows the mean transition time parametric image on a basal SA slice. The corresponding DE image with the superimposed myocardial contours is shown in Fig. 5(c). For this slice, the expert contraction scores are H-N-N-N-H-AD for the segments 1 to 6. The segment 6, classified as AD, appears in blue on the $A_v(x, y)$ image (negative values), corresponding to an inverse movement and in green on the mean transition time image, corresponding to a delayed contraction. The widths of the colored bands are also reduced on the amplitude image in the segments 1 and 5, classified as H. The DE-CMR image confirms the presence of transmural scar in the segment 6.

The amplitude to time ratio was automatically obtained using the PAMM images and the tele-diastolic endocardial contours on all the 348 myocardial segments. An analysis of variance was achieved to compare the ATR values with the visual contraction study. Significant differences were found between AD and H segments ($p < 0.003$) and between H and N segments ($p < 0.0001$).

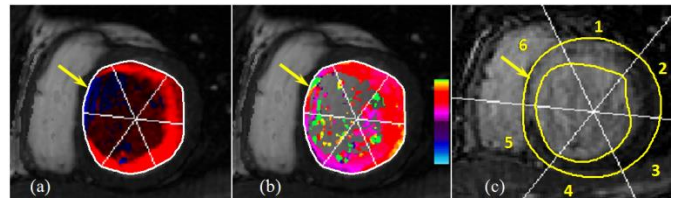

Fig. 5. Combined analysis of regional wall motion analysis using PAMM and MIE: (a) parametric image $A_b(x, y)$ in gray levels and superimposed parametric image $A_v(x, y)$ in colors (red represents positive values, blue negative values), (b) parametric image of mean transition time in "rainbow" color ranging from 0 to 0.5, (c) corresponding DE image with the superimposed myocardial contours and the myocardial subdivision into 6 segments.

*E. Relationship between MIE and Contraction Function*

For the statistical analysis of the results and to allow the segment-to-segment comparison between enhancement and contraction, the automatic and the visual transmurality indexes were averaged on the 3 sub-segments of each myocardial segment, for all the studied patients.

Fig. 6 illustrates the results of the segment-to-segment comparison between the visual (expert) and automatic MIE scores and the visually assessed contraction function, respectively shown on the left and on the right of Fig. 6. We can see that the expert mean MIE scores are higher in regions classified by the visual myocardial function analysis as H or AD than in regions classified as normal, where the hyperenhancement is low or inexistent. Fig. 6 also shows that the automatically obtained mean MIE scores are similarly analyzed compared to the contraction function. For both expert and automatic MIE scores, a significant difference was observed between normal and pathological segments but there was no significant difference allowing discriminating H and AD segments.

Moreover, the estimated ATR values were compared to the MIE quantification scores obtained by the visual expertise and by the automatic method. Fig. 7 shows the ATR values estimated using PAMM on the 348 segments as a function of the mean values of the MIE scores in segments classified as NDE (No Delayed Enhancement) and DE (Delayed Enhancement) by the expert (left) and by the automatic quantification (right).

Finally, we observed that the ATR significantly decreases in pathological myocardial segments, yielding a clear separation between normal (NDE) and pathological (DE) segments. The high correlation between the automatically deduced ATR parameter and the visual assessment of contraction thus confirms the reliability of ATR and constitutes a validation of the automatic assessment.

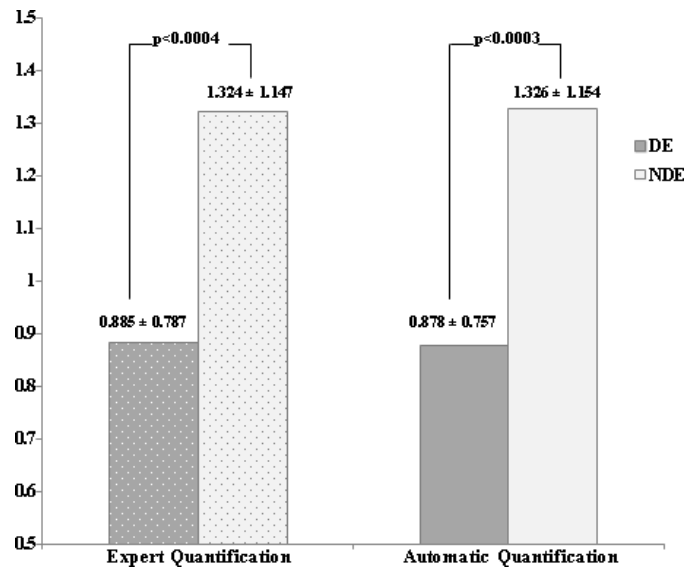

Fig. 6. Relationship between the MIE and the regional wall motion in normal (N), hypokinetic (H) and a/dyskinetic (AD) segments. The mean values of the visual MIE scores obtained by the expert are shown on the left and the mean values of the automatic MIE scores, obtained using the Fuzzy c-means classification are shown on the right (****: $p<0.0001$, NS: non-significant).

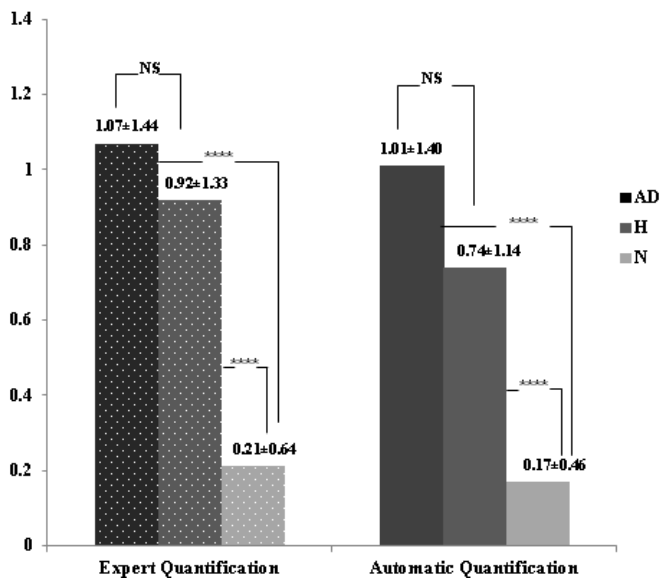

Fig. 7. Amplitude to Time Ratio (ATR) parameter on the 348 segments, as a function of the enhancement (DE or NDE) assessed by the visual expert (left) and by the automatic method (right). A significant decrease in the ATR values in pathological myocardial segments yields a clear separation between the segments classified as normal (NDE) and pathological (DE) both with the automatic and visual quantification scores.

## VI. CONCLUSION AND FUTURE WORK

The extraction of quantitative parameters from CMRI for the accurate diagnosis and follow-up of heart diseases requires segmentation and registration procedures. The use of computer-assisted methods for the detection and delineation of the heart structures and injured zones allows fusing anatomical and functional information, reducing variability, and saving computing time in the clinical practice.

The first objective of this study was to propose a software integrating computer-assisted methods and allowing the estimation of functional cardiac parameters. The second goal was to determine the relationship between the regional hyperenhancement and the associated regional myocardial contraction. With this aim in view, we proposed a complete procedure for CMRI data processing, including generic tools for the semi-automated segmentation of the LV myocardial contours and two steps coarse then refined 3D rigid registration method using the NMI as similarity criterion to correct for motion on the DE/Cine cardiac acquisitions. This two-step procedure showed high DE/Cine registration accuracy and allowed a considerable refining of global spatiotemporal alignments.

Dedicated tools were then developed to quantify the MIE and to study the regional function from DE and Cine CMRI. The whole process was successfully applied to 30 DE/Cine datasets and the experiments demonstrated good accuracy and robustness to user intervention.

The registration results showed high quality alignment of all the processed data and allowed improving considerably the global spatiotemporal alignment. This was confirmed by the comparison of the qualitative and quantitative evaluation with the expert visual scores, considered as the reference. A high agreement of 91% between the automatic MIE quantification and the visual expert evaluation was observed. The 9% rate of differences between the automatic and visual assessment can be explained by several factors such as the difficulty to visually define the segment limits, registration and segmentation inaccuracies, and noise artifacts.

One of the limitations of the presented procedure is that the segmentation process is time-consuming since epicardial and endocardial contours are segmented separately, slice after slice. Moreover, the user may have to test several sets of parameters to reach the best myocardial segmentation. Nevertheless, the semi-automatic method that we propose has precision performances close to that of manual methods [34]-[37]. Furthermore, authorizing a minimal number of user interventions allows overcoming the difficulties involved in the segmentation. The fully automatic definition of the cardiac ROI and a 3D segmentation, which constitute a part of our current work, will reduce the recourse to user intervention.

Using this process, we finally addressed the problem of defining the relationship between the myocardial enhancement and the associated regional contraction. In order to quantify the regional contraction from the Cine-CMRI using PAMM, a new index combining the amplitude and mean transition time indexes is proposed. This index, defined as the Amplitude to Time Ratio (ATR), decreases as the severity of the contraction dysfunction increases and allows discriminating with high significance not only between normal and hypokinetic segments, but also between akinetic/dyskinetic and hypokinetic segments. The comparison of the automatically assessed MIE scores and the estimated ATR showed that the myocardial contraction is highly related to the sub-endocardial enhancement. We thus confirmed that the segmental myocardial dysfunction increases with the extent of the myocardial enhancement.

## ACKNOWLEDGMENTS

The authors wish to thank Dr. Alban Redheuil for his second reading of clinical data.